\documentclass[a4paper,twoside]{article}
\newcommand{\affil}[1]{$^{\rm #1}$}
\usepackage[authoryear]{natbib}
\bibpunct{(}{)}{;}{a}{}{,}
\usepackage{graphicx}
\date{} 
\newcommand{\epm}{e^{\pm}}

\title{\large\bf\flushleft Search for Synchrotron Emission from Secondary Leptons in Dense Cold Starless Cores.}
\author{\parbox{\textwidth}{\flushleft
\vspace{-0.5cm}
{\it D.I. Jones\affil{A,B,D}, R.J. Protheroe\affil{A} and R.M. Crocker\affil{A,C}}\\
\vspace{0.4cm}
{\small \affil{A}\,School of Chemistry \& Physics, University of Adelaide, Adelaide, South Australia, 5000}\\
{\small \affil{B}\,Australia Telescope National Facility, CSIRO, P.O. BOX 76 Epping, NSW 1710, Australia}\\
{\small \affil{C}\,School of Physics, Monash University, Clayton, Victoria 3168, Australia} \\
{\small \affil{D}\,Email: djones@physics.adelaide.edu.au}}}
\begin{document}
\twocolumn[
\begin{changemargin}{.8cm}{.5cm}
\begin{minipage}{.9\textwidth}
\vspace{-1cm}
\maketitle
%
%
\small{\bf Abstract: We report radio continuum observations with the Australia Telescope Compact Array of two molecular clouds. The impetus for these observations is a search for synchrotron radiation by cosmic ray secondary electrons/positrons in a region of enhanced density and possibly high magnetic field.  We present modelling which shows that there should be an appreciable flux of synchrotron above the more diffuse, galactic synchrotron background.  

The starless core G333.125-0.562 and infrared source IRAS 15596-5301 were observed at 1384 and 2368 MHz.  For G333.125-0.562, we find no significant levels of radio emission from this source at either frequency, nor any appreciable polarisation: we place an upper limit on the radio continuum flux from this source of 0.5 mJy beam$^{-1}$ at both 1384 and 2368 MHz.  Due to the higher than expected flux density limits, we also obtained archival ATCA data at 8640 MHz for this cloud and place an upper limit on the flux density of 50 $\mu$Jy beam$^{-1}$.  Assuming the cosmic ray spectrum is similar to that near the Sun, and given the cloud's molecular density and mass, we place an upper limit on the magnetic field of 500 $\mu$G.  

IRAS 15596-5301, with an RMS of 50 $\mu$Jy beam$^{-1}$ at 1384 MHz, shows an HII region consistent with optically thin free-free emission already detected at 4800 MHz.  We use the same prescription as G333 to constrain the magnetic field from this cloud to be less than 500 $\mu$G.  We find that these values are not inconsistent with the view that magnetic field values scale with the average density of the molecular cloud. }

\medskip{\bf Keywords:} synchrotron radiation, cosmic rays, molecular clouds, starless cores:
  general --- molecular clouds: individual: G333.125-0.562, IRAS15596-5301

\medskip
\medskip
\end{minipage}
\end{changemargin}
]
\small

\section{Introduction}
The Galactic synchrotron emission is usually thought to be due entirely to primary cosmic ray electrons and electron/interstellar medium (ISM) interactions.  However, a significant contribution could also be made by electrons and positrons (hereafter $\epm$) produced as secondaries by interactions of cosmic ray (CR) nuclei with atomic nuclei, in molecular clouds where conditions (such as magnetic field strength and density) could be favourable for the production of synchrotron photons from the secondary $\epm$.  

CRs play an important role in molecular cloud evolution by partially ionizing even the cold molecular gas, and thereby affecting the dynamics of cloud collapse by coupling the magnetic field to
the partially ionized gas.  This in turn could result in amplification of the ambient magnetic field during cloud collapse, and give rise to a correlation between density of molecular gas and magnetic field.  \citet{Crutcher1999} has found just such a relation between magnetic field and average H$_2$ number density, $n_{H_2}$ (cm$^{-3}$), of individual molecular clouds,
\begin{equation}
B = 0.1 \left({n_{H_2} \over 10^4\mbox{ cm}^{-3}}\right)^{0.47} \mbox{~~~mG}.
\end{equation}

We recently conducted a radio continuum study with the Australia Telescope Compact Array (ATCA)\footnote{The Australia Telescope Compact Array is a part of the Australia Telescope which is funded by the Commonwealth of Australia for operation as a National Facility managed by CSIRO}, to search for evidence of synchrotron emission due to secondary $e^{\pm}$ in molecular clouds.  As will be shown, the expected emission is linearly proportional to target mass, and should have a negative spectral index $\alpha\sim -0.7$ (for $S_\nu\propto\nu^\alpha$, where $\alpha$ is the spectral index).  We therefore selected massive clouds such as the Galactic Centre (GC) giant molecular cloud (GMC), Sagittarius~B2 (Sgr~B2) because of the higher expected CR density in the region.  Because of the considerably more complex structure of this cloud, the results of the observations of Sgr B2 will be presented elsewhere (Jones et~al., in preparation; see also Crocker, et al. 2007).  We also selected nearby ($\sim$ 3-4~kpc) cores of giant molecular cloud (GMC) complexes, and in order to maximise the probability of detection, we selected clouds which have little or no known thermal radio emission.  The sources selected were the infrared source, IRAS~15596-5301 (hereafter IRAS15596) \citep{Garay2004}, for its low levels of radio continuum emission at 4800 MHz and no spectral information, and G333.125-0.562 (hereafter G333) \citep{Garay2002}, because of its status as a starless core and lack of radio continuum observations.  

\begin{table}[h]
	\begin{center}       
		\caption{Distances to and physical parameters for the two clouds.} \label{cloudParameters}	
	\begin{tabular}{lcc}
	\hline& G333$^a$  & IRAS15596$^b$  \\
	\hline Mass (M$_\odot$) & 2.3$\times10^3$$^{,c}$  & 1.4$\times10^3$  \\
	$n_{H_2}$ (cm$^{-3}$) & $2\times10^5$ & $4\times10^5$ \\
	$T$ (K) & 13$^c$ & $\sim27$ \\
	$d$ (kpc) &3.6 & 4.6 \\
	$R$ (pc) & 0.34 & 0.47\\
       	\hline
	\end{tabular}	
	\medskip\\
	$^a$ \citet{Garay2004}; $^b$\citet{Garay2002}; $^c$ \citet{Lo2007}\\
	\end{center}
	
\end{table}
The densities of the two clouds, are listed in Table \ref{cloudParameters}, and using the Crutcher prescription, we expect magnetic field values of $400^{+120}_{-80}$ and $550^{+200}_{-140}$ $\mu$G for G333 and IRAS15596 respectively.  The error for the expected values is derived from the $\pm0.08$ uncertainty of the exponent quoted by \cite{Crutcher1999}.

The production rate of secondary electrons and
positrons depends only on the spectrum and intensity of cosmic ray nuclei, and the density of the interstellar matter. We use the production rate of electrons and positrons $q_m(E)$, per solar mass of interstellar matter per unit energy (M$_\odot^{-1}$~GeV$^{-1}$~s$^{-1}$), for the cosmic ray spectrum and composition observed above the Earth's atmosphere based on Figure~4 of \citet{Moskalengko1998} (we note here that the choice of CR spectral index and normalisation is very important, since the clouds are so close, there is no reason to suppose that the spectrum should deviate significantly from that observed above our atmosphere).  The production spectrum of electrons and positrons per unit volume per unit energy at position $\vec{r}$ is then the product of the density of interstellar gas at position $\vec{r}$ multiplied by the production rate per unit mass ($e^\pm$ cm$^{-3}$ GeV$^{-1}$ s$^{-1}$)
\begin{equation}
q(E, {\bf r})=n_{H_2}({\bf r})q_m(E)2m_{H}.
\end{equation}

For moderate molecular cloud densities $n_H>10^2$~cm$^{-3}$ and magnetic fields $B>10^{-5}$ G, the relatively short energy loss time justifies neglecting diffusive transport of electrons \emph{within the cloud}.  Additionally, we note that the small radii of the clouds themselves, should not prevent the $\sim$ GeV CR protons from freely entering the clouds.  Hence we assume that the clouds are `bathed' in GeV CRs, and the charged secondary particles produced decay and cool within the confines of the cloud. 

Having obtained the production spectrum, one then readily obtains by numerical integration the ambient number density of electrons and positrons, per unit energy, $n^\pm(E,{\bf r})(e^\pm$ cm$^{-3}$ GeV$^{-1}$), at various positions $\vec{r}$ within the molecular cloud complex: 
\begin{equation}
n(E, {\bf r}) = {\int_E^\infty q(E, {\bf r}) dE \over dE/dt},
\end{equation}
where $dE/dt$ is the total rate of energy loss of electrons at energy $E$ due to ionization, 
bremmstrahlung and synchrotron emission (because of the energies involved, we neglect 
positron annihilation, and assume electrons and positrons suffer identical energy losses). 
Electrons lose energy by ionization losses in neutral molecular hydrogen at a rate (in GeV s$^{-1}$)
\begin{equation}
{dE\over dt}_{\rm ioniz}=-5.5 \times 10^{-17}  \left({n_{H_2}\over {\rm 1 \; cm^{-3}}}\right) \times (\ln \gamma+6.85),
\end{equation}
\begin{equation}
{dE\over dt}_{\rm bremss}=-1.5 \times 10^{-15}\left({E\over {\rm 1 \; GeV}}\right) \times \left({n_{H_2}\over {\rm 1 \; cm^{-3}}}\right).
\end{equation}
\begin{equation}
{dE\over dt}_{\rm synch}=-1.0\times 10^{-12} \times \left({B_\perp \over {\rm 1 \; gauss}}\right)^2 \times \gamma^2,
\end{equation}
where $B_\perp$ is the component of magnetic field perpendicular to the electron's direction. For an isotropic electron population $\langle B_\perp\rangle= 0.78B$.  The synchrotron emissivity, $j_\nu$ (in erg cm$^{-3}$ s$^{-1}$ sr$^{-1}$ Hz$^{-1}$), is then calculated using standard formulae in synchrotron radiation theory \citep{RybickiLightman1979}
\begin{eqnarray}
j_\nu={\sqrt{3} \; e^3 \over 4\pi m_ec^2} \left({B_\perp\over {\rm\;1\;Gauss}}\right)\int_{m_ec^2}^\infty F(\nu/\nu_c)n(E, {\bf r})dE, \\
\nu_c =  4.19\times 10^6 (E / m_ec^2)^2 (\frac{B_\perp}{\rm \; 1 \; Gauss})  \mbox{ ~~~ Hz}, \\
 e = 4.8\times 10^{-10}  \mbox{ ~~~ esu},\\
m_ec^2 = 8.18\times 10^{-7}  \mbox{ ~~~ erg},,\\
F(x) = x\int_x^\infty K_{{5 \over 3}}(\xi)d\xi.
\end{eqnarray}
where $K_{\frac{5}{3}}(x)$ is the modified Bessel function of order 5/3. 

\begin{figure}[h]
\centering
\includegraphics[scale=0.45,angle=0]{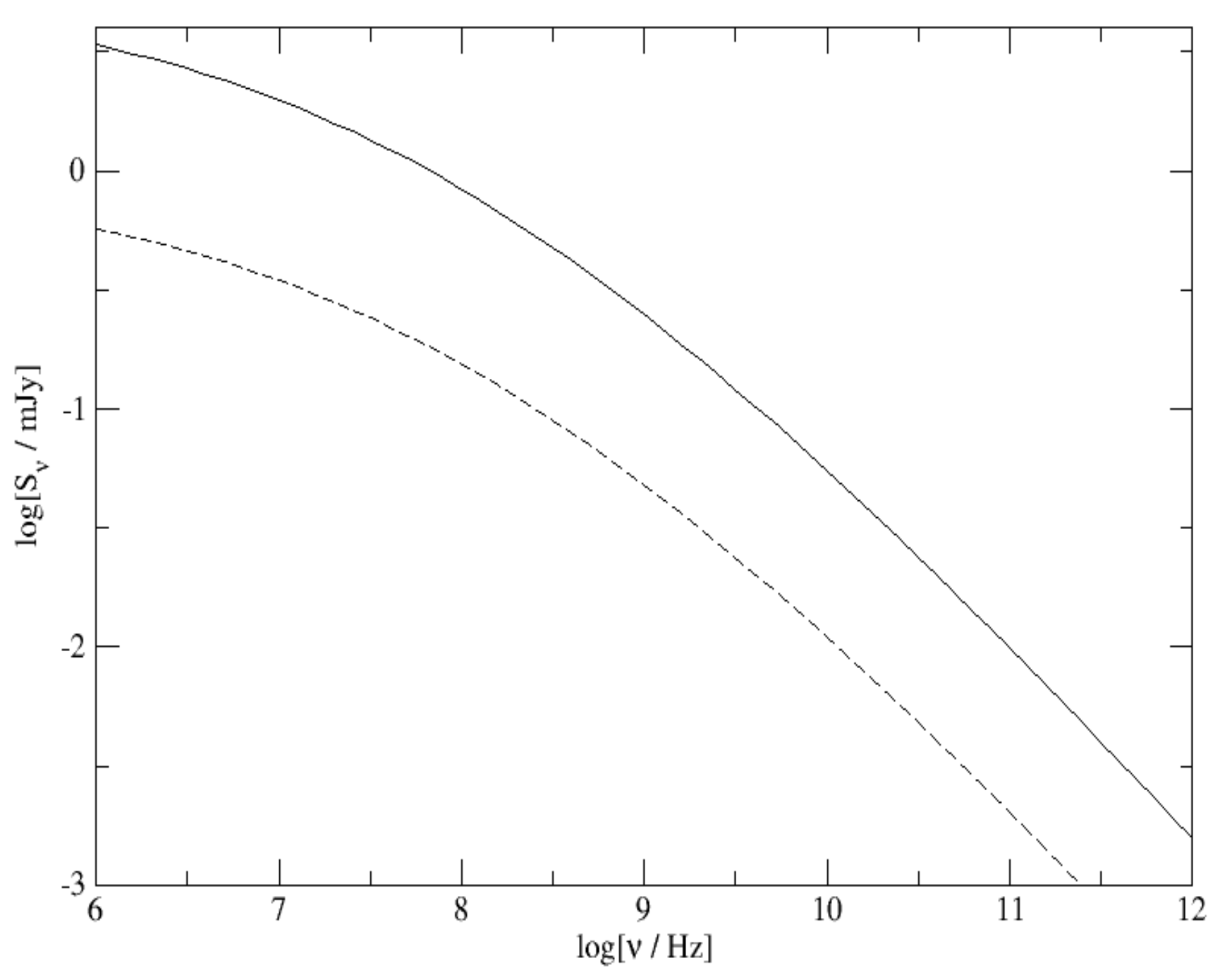}
\caption{Log-log plot of the expected flux from the two cores chosen for the observations; solid line is G333, dotted line is IRAS15596.  The parameters used are listed in Table 1, along with the value of the magnetic field as predicted by the \cite{Crutcher1999} scaling relation.  Additionally, the local CR spectrum is assumed (for reasons, see text).}
\label{fluxes}
\end{figure}

Figure \ref{fluxes} shows the expected synchrotron emission from the two clouds using the above outlined method, the physical parameters listed in Table 1, and the Crutcher scaling relation.  This shows that a sensitivity on the order of 100 $\mu$Jy (at centimeter wavelengths) should provide a robust detection of these clouds at centimeter wavelengths.

\section{Observations}\label{observationsSection}
\subsection{Observations \& Data Reduction}
Single pointing ATCA radio continuum observations were made on 2005 December $4-5^{th}$, using the 1.5C configuration utilising five of a possible six antennae and covering east-west baselines from 76.5 meters to 1.4 km.  Observations were made simultaneously at the frequencies 1384 MHz ($\lambda=20$ cm) and 2368 MHz ($\lambda=13$ cm), each spanning 128 MHz in bandwidth and recording all four polarisation products.  At these frequencies, the FWHM of the ATCA beam is 34.5$'$ and 20.1$'$ at 1384 and 2368 MHz respectively.  The total integration time for each source was 720 minutes obtained from 50 minute scans over 2 hours on the 4$^{th}$ for G333.125-0.562 and 2 hours on the 5$^{th}$ for IRAS 15596-5302. The remaining observations were gained by using 15 minute cycles over 10 hours on the 5$^{th}$, in order to provide excellent (\emph{u,v})-coverage.  The calibrator PKS 1613-586 was observed before and after every on-source scan in order to correct the amplitude and phase of the interferometer data for atmospheric and instrumental effects.  The flux density was calibrated by observing PKS 1934-638 (3C84), for which the flux densities of 14.94 and 11.59 Jy at 1384 and 2368 MHz respectively were adopted.  Due to PKS 1934-638 being below the horizon, the array was system calibrated using PKS 0823-500, and PKS 1934-638 was observed at the end of both of the observing runs, and the flux densities for both PKS 0823-500 and PKS 1613-586 were bootstrapped from PKS 1934-638. 

Data reduction was performed using the MIRIAD software package using standard calibration procedures, except as noted above.  Imaging was then performed by Fourier transforming the interferometer data using the MIRIAD task \emph{invert}.  The images were then CLEANed using the task \emph{maxen}, where the flux was constrained using a total flux estimate gained from a CLEANed image gained using the task \emph{clean}.  The synthesised (FWHM) beams were at 1384 and 2368 MHz respectively: $20''.3\times18''.7$ and $11''.3\times9''.9$ for G333 and $19''\times10''.5$ and $15''\times9''.5$ for IRAS15596.  The noise in the images were determined by adding in quadrature, the standard deviation of the flux of the non-primary-beam-corrected image, integrated within a number of regions roughly corresponding to  the beam size in areas of the image which lack an obvious source.  The RMS noise was determined to be 0.5 mJy beam$^{-1}$ at both 1384 and 2368 MHz for the G333 cold core, and 30 and 50 $\mu$Jy beam$^{-1}$ at 1384 and 2368 MHz respectively for IRAS15596.  The RMS for the IRAS source are close to the theoretical noise limit of the telescope, whereas for the G333 cold core, it is $\sim10$ times the theoretical noise limit due to strong emission nearby.

\subsubsection{Additional Data}
In order to better characterise any emission from the chosen clouds, we have obtained additional, archival data:  843 MHz Molongolo Observatory Synthesis Telescope (MOST) data for both G333 and IRAS15596; and archival 8640 MHz ATCA data for G333.

843 MHz MOST data was obtained from the Sydney University Molongolo Synthesis Survey (SUMSS) catalogue \citep{SUMSS}.  SUMSS has a synthesised beam of 43$''\times43''\sin^{-1}(\delta)$, where $\delta$ is the declination of the source, and a limiting peak brightness of between 6 and 10 mJy beam$^{-1}$, depending on source declination \citep{SUMSS}.

We also obtained archival ATCA data for the G333 cloud for two reasons:  firstly to check for low level, optically thick emission at higher frequencies, and secondly (as will be show in \S4) the RMS noise limit for the lower frequencies are much higher than the expected emission shown in Figure \ref{fluxes}.  The data was from Project Code C483, taken in 1995\footnote{It can be accessed from the ATCA on-line archive: http://atoa.atnf.csiro.au/}.  It used the 6A and C configurations of the ATCA, utilising all 6 possible antennae, with baselines from 153 and 337 m to 6 km, for the 6C and 6A configurations respectively.  The observations were centered at a frequency of 8640 MHz ($\lambda=3$ cm) spanning 128 MHz in bandwidth (4800 MHz ($\lambda=6$ cm) observations were taken simultaneously, however, these were in spectral line mode, so were not used).  It is unclear how many polarisation products were recorded, however only Stokes I is used here.

These data were reduced using the steps outlined above in \emph{miriad}, using PKS 1934-638 as the primary calibrator (assuming the same flux densities noted above), and PKS 1600-40 as the secondary calibrator (the secondary calibrator was chosen by plotting the $(u,v)$-distance as a function of time for the calibrators and source field, and selecting that which had observations of the caibrator interspersed).  At 8640 MHz, the primary beam is $5.5'$ and the resulting synthesised beam is $1.2''\times0.8''$ at a position angle of 24$^\circ$.  The 1-$\sigma$ RMS noise for the image was found to be $\sim50$ $\mu$Jy beam$^{-1}$, just above the expected noise level.

\section{Results}

\subsection{G333.125-0.562}
\begin{figure}[h]
\centering
\includegraphics[scale=0.28,angle=0]{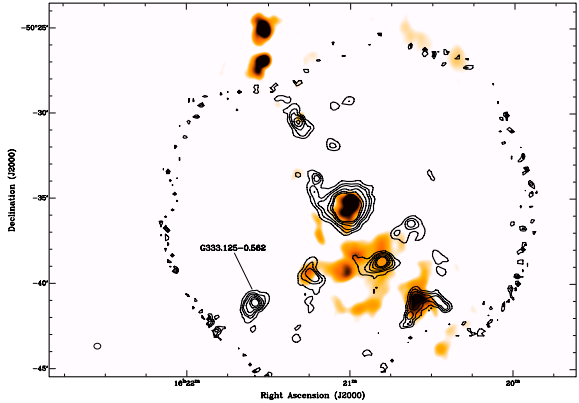}
\caption{G333.125-0.562: ATCA 1384 MHz total intensity image overlaid with (black) 1.2 mm (Garay, 2004) contours at 0.18, 0.36, 0.54, 0.90, 1.36 and 1.80 Jy beam$^{-1}$.  The intensity image uses a logarithmic transfer function - $\log(val-min)$ - to bring out emission on all scales.  The beam for the 1384 MHz emission is located in the lower left-hand corner.  The G333 cold core is labelled.}
\label{G333}
\end{figure}

Figure \ref{G333} shows the total intensity image of the region around the G333 cold core from the 1384 MHz ATCA observations, overlaid with the 1.2 mm contours as described in \citet{Garay2004}.  The bright emission in the image is a large ridge of emission known as the RCW106 giant molecular cloud (GMC).  It is obvious that even with a logarithmic transfer function applied to the image, there is no source coincident with the 1.2 mm dust emission from G333 as observed by \citet{Garay2004} down to the 1-$\sigma$ level at 843, 1384 or 2368 MHz.  There was no significant polarised emission detected at either 1384 or 2368 MHz.

Since the noise level from the 1384 and 2368 MHz images does not match the expected source strength, due to the extended emission in the beam, we imaged the region using the archival 8640 MHz ATCA data described above.  The 8640 MHz data, whilst at a frequency where it could be expected to be confused with thermal emission, contains a smaller primary beam (5$'$ as opposed to $\sim30'$ at 1384 MHz) and an RMS noise limit below the expected flux shown in Figure \ref{fluxes}.  The image does not contain any significant continuum emission, and hence is not shown here.

\subsection{IRAS 15596-5301}
\begin{figure}[h]
\centering
\includegraphics[scale=0.3,angle=0]{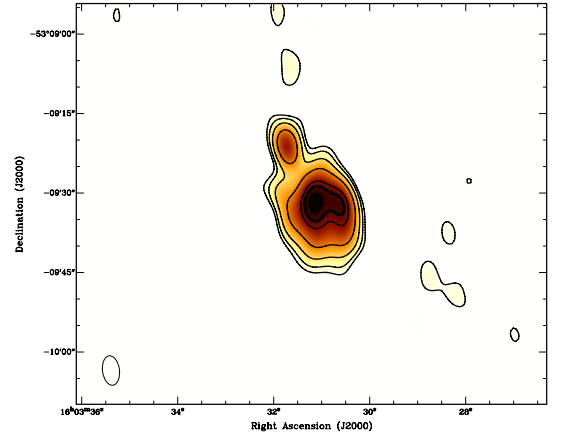}
\caption{IRAS15596-5302: ATCA 2368 MHz total intensity image overlaid with (black) 2385 MHz contours at 0.4 0.8 1.6 3.2 6.4 8.0 and 10.0 mJy beam$^{-1}$ with the beam in the lower left-hand corner.  The same transfer function as Figure \ref{G333} image has been applied.}
\label{IRAS15596}
\end{figure}

Figure \ref{IRAS15596} shows the total intensity emission at 2368 MHz overlaid with 2368 MHz ATCA contours.  There is a source coincident with the 4800 MHz source from \citet{Garay2002}.  The peak fluxes for this source are 19.6 and 14 mJy beam$^{-1}$ at 1384 and 2368 MHz respectively.  Integrating within a region corresponding to the 5-$\sigma$ level at each frequency, the total integrated fluxes for this source are 59, 53.7 and 49 mJy at 843, 1384 and 2368 MHz respectively.   

\section{Discussion}
\subsection{G333.125-0.562}
\begin{figure}[h]
\centering
\includegraphics[scale=0.3,angle=0]{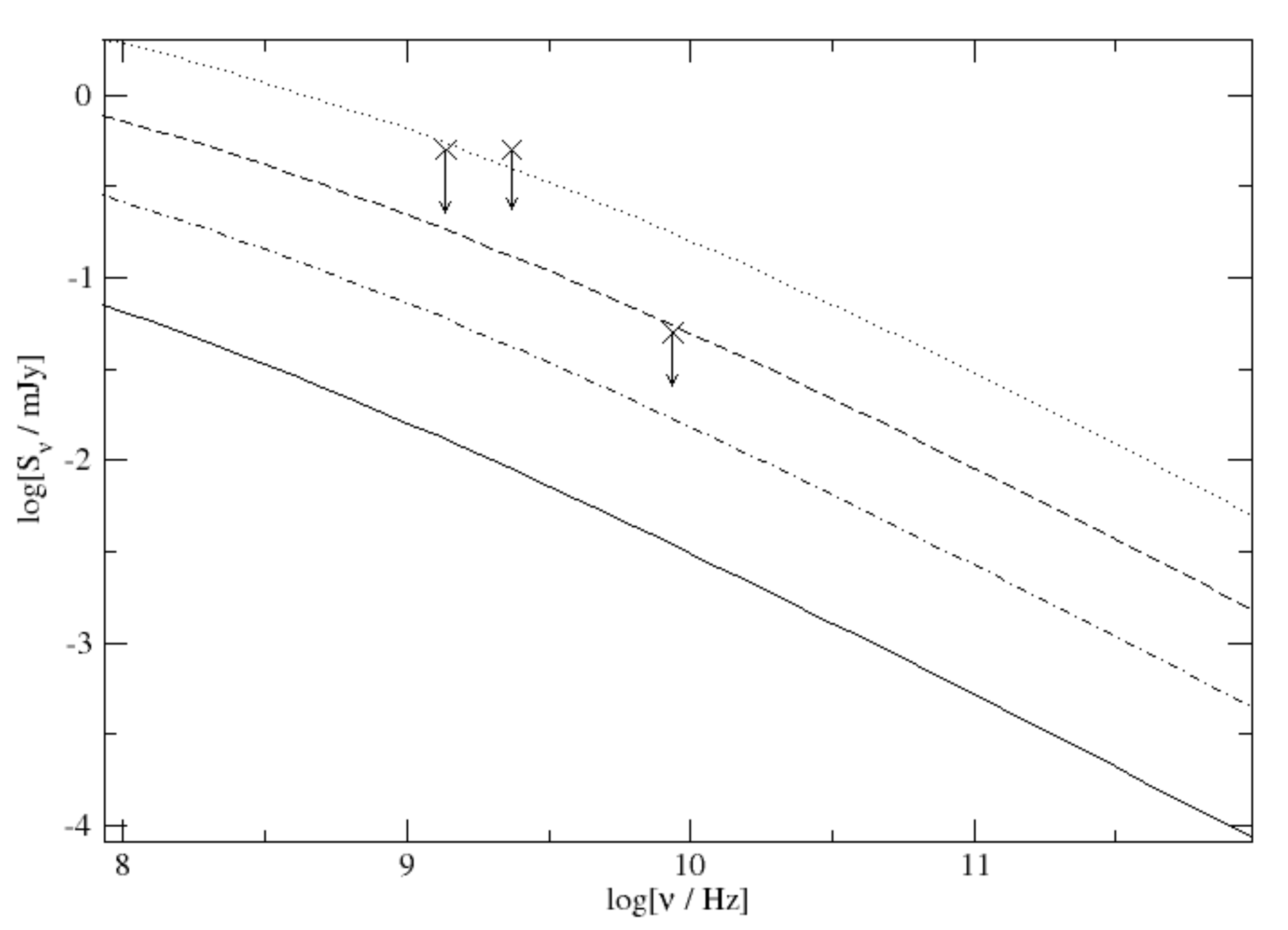}
\caption{Predicted flux from the G333 cold core for different magnetic field strengths: 100 $\mu$G (solid), 250 $\mu$G (dot-dash), 500 $\mu$G (dash), and 1 mG (dotted).  The crosses are (from left) the 1-$\sigma$ noise levels of 0.5 mJy, 0.5 mJy and $\sim50$ $\mu$Jy beam$^{-1}$ derived from the 1384, 2368 and 8640 MHz datum respectively.  The physical parameters used in the determination of the flux densities are listed in Table 1.}
\label{G333FluxLimit}
\end{figure}

Figure \ref{G333FluxLimit} shows the expected synchrotron emission assuming a local CR spectrum, for a variety of magnetic field strengths.  It can clearly be seen that a field of 1 mG would provide a 5-$\sigma$ detection, whilst for a field strength of 500 $\mu$G we predict a flux of $\sim$0.5 mJy at 1 GHz. Additionally, assuming no thermal contamination, a 500 $\mu$G field strength would produce a probable detection at 8640 MHz.

Since there is no emission at 843, 1384, 2368 or 8640 MHz down to the 1-$\sigma$ noise level of 0.5 mJy beam$^{-1}$ at 1384 and 2368 MHz; 1 mJy beam$^{-1}$ at 843 MHz; and 50 $\mu$Jy beam$^{-1}$, we place an upper limit on the magnetic field of 500 $\mu$G.  Despite being lower in frequency (hence having the expectation of a higher synchrotron emissivity), the upper limit on the magnetic field is derived from the archival ATCA data, since the SUMSS catalogue has a limiting peak brightness for sources $\delta>-50^\circ$ is 6 mJy beam$^{-1}$ \citep{SUMSS}: an order of magnitude worse than the limits derived from our images and even worse still that the archival data at 8640 MHz.  

\subsection{IRAS 15596-5301}
\begin{figure}[h]
\centering
\includegraphics[scale=0.4,angle=0]{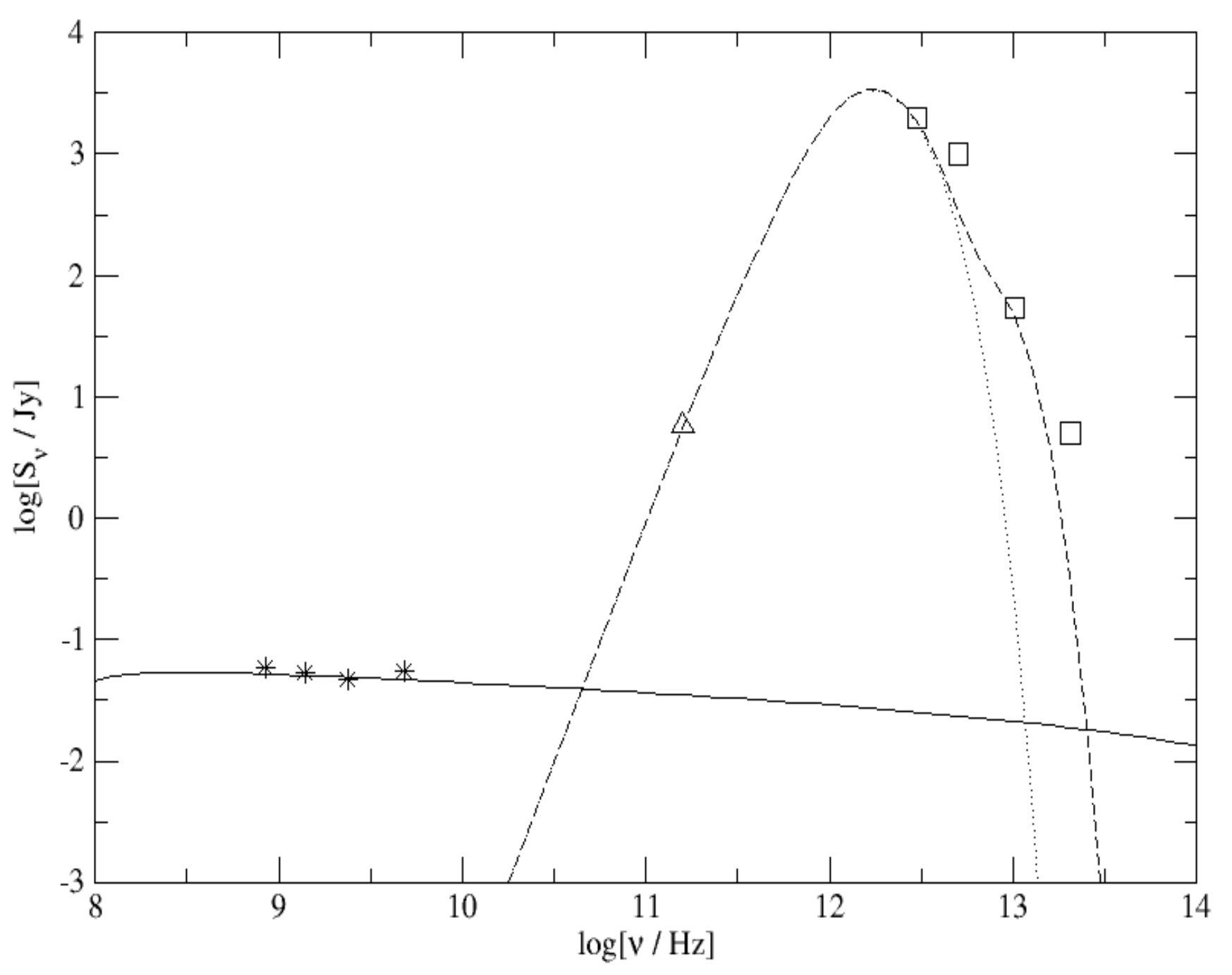}
\caption{Spectral Energy Distribution (SED) for IRAS 15596.  The data points are: stars represent the 843, 1384, 2368 and 4800 MHz SUMSS and ATCA data (integrated as described in the text); triangle is the 1.2 mm emission from \citep{Garay2002}; the squares are the IRAS fluxes as reported in \citep{Garay2002}.  The fluxes modelled are the free-free emission (solid curve) using emission measures as described in \citep{Garay2002}, and electron temperature of 10,000 K.  The modified blackbody curves are described in \S5.2 and are for dust temperatures of 27 (dash) and 100 K (dot-dash), using the parameter set in \citet{Garay2002}.}
\label{IRAS15596FluxLimit}
\end{figure}

Figure \ref{IRAS15596FluxLimit} shows the full spectrum of IRAS15596.  Two components are modelled: free-free emission, tracing the emission from the HII regions; and a modified Blackbody curve, following the dust and infra-red emission.  Following \citet{Garay2002}, the free-free emission was modelled using an emission measure ($EM\approx n_ed$ pc cm$^{-6}$), for an electronic temperature of 10,000 K.  The higher frequency emission was fitted using two modified blackbody curves, of the form:
\begin{equation}
  S_\nu=B_\nu(T_d)[1-\exp(-\tau_\nu)]\Omega_s
\end{equation}
where $B_\nu(T_d)$ is the Planck function, $T_d$ is the dust temperature, and $\Omega_s$ is the solid angle subtended by the dust emitting region (see figure \ref{IRAS15596}).  The optical depth, $\tau_\nu$, is allowed to vary with frequency; $\tau_\nu=(\nu/\nu_0)^\beta$, where $\nu_0$ is the frequency at which the optical depth is unity, and $\beta=2$, consistent with high mass star forming regions \citep{Garay2002}.  The parameters used to model the dust emission can be found in \citet{Garay2002}.

From the integrated fluxes presented in the previous section, it can be seen that the radio continuum emission traces the optically thin, free-free emission well with a spectral index between all radio continuum frequencies (843 to 4800 MHz) of $\alpha\sim-0.1$.  We have placed upper limits on the magnetic field in this cloud by turning the original question around, that is: \emph{how high can the magnetic field be without significantly altering the optically thin free-free emission spectrum that is observed?}.  Placing a conservative 10\% characteristic error on the fluxes (due to RMS noise of the image, plus systematical uncertainties), a magnetic field is implied to be of the order of 0.5 mG is required to produce $\sim$5-6 mJy of flux at 1 GHz.  Taking 0.5 mG as the upper limit, this is not inconsistent with the scaling relation from \citep{Crutcher1999}.

\section{Conclusion}
We undertook a search for synchrotron emission from molecular clouds due to production of secondary $e^{\pm}$.  We have modelled the expected flux using the local cosmic ray spectrum, and shown that the emission should be observable above the more smoothly distributed, diffuse synchrotron emission of the galaxy if the magnetic fields in the clouds are sufficiently high.  We had initially selected three clouds, Sgr B2 (Jones, \emph{et al.}, in preparation) because of the expected increase in cosmic ray density near the GC, and two other clouds because they are isolated, dense, cold, dark clouds, with little evidence of star formation, and little or no previously detected radio emission.  Additionally, their smaller distance makes the assumption of a local CR spectrum safer.  For these nearby clouds, our conclusions can be summarised as follows:
\begin{enumerate}
\item G333.125-0.562, from 1.2mm, MSX and IRAS observations, was designated as a starless core \citep{Garay2004}.  We have presented the first high resolution, radio continuum observations of this core, and do not find any radio emission at any frequency.  This is in line with previous determinations that this cloud is not observed at any wavelength longer than 1.2 mm \citep{Lo2007}.  We place a 1-$\sigma$ upper limit on the flux from this cloud of 500 $\mu$Jy beam$^{-1}$ at 1384 and 2368 MHz, and 50 $\mu$Jy beam$^{-1}$ at 8640 MHz.  The flux density upper limits in turn, were used to place an upper limit on the magnetic field of this cloud of 500 $\mu$G, consistent with the magnetic field - density correlation found by \citep{Crutcher1999}. 

\item  From 4800 MHz observations taken with ATCA, \citep{Garay2002} concluded IRAS 15596-5301 to be a molecular cloud undergoing massive star formation.  The flux at lower frequencies (including the 843 MHZ SUMSS data), presented here, confirms this.  The spectral index is consistent with free-free emission from an HII region - indicating that this cloud is in an advanced stage of star formation.  Additionally, we constrain the magnetic field to be below 0.5 mG, a value which is also consistent with the scaling relation found in \citet{Crutcher1999}.

\item  Though we failed to detect synchrotron emission from either cloud, we note that the initiating particles are of $\sim$GeV energies.  If there are sufficient numbers, as our production spectrum indicates, then the upcoming GLAST mission may be able to probe the distribution of the CR spectum in our galaxy by selecting a sample of suitable clouds to search for $\gamma$-ray emission from $\pi^0$ decay.  
\end{enumerate}

We conclude that we did not find any evidence for synchrotron emission due to $e^{\pm}$ in these clouds.  We note, however, that thermal emission from already formed massive stars, and nearby thermal emission tends to confuse possible detections, due to the low levels expected from the synchrotron emission.



\section*{Acknowledgments} 
We would particularly like to acknowledge the help and advice of Ron Ekers in the observations and preparation of the data and as an active participant in the early stage of this research.  We would also like to acknowledge and thank Anne Green and colleagues for the use of their 843 MHz SUMSS data. DIJ thanks Kate Brooks and Guido Garay for the use of their data and help during the preparation of this paper, and Gavin Rowell for discussions on the CR spectrum.  This research has made use of the SIMBAD database, operated at CDS, Strasbourg, France.

\end{document}